\documentclass[12pt, a4paper]{article}

\usepackage[latin1]{inputenc}
\usepackage{graphicx}
\usepackage{url}
\usepackage{amsmath}
\usepackage{amssymb}
\usepackage{ifpdf}

\newcommand{\titleman}[1]{\begin{center}{\bf \LARGE
#1}\end{center}}

\newcommand{\bm}[1]{\mbox{\boldmath{$#1$}}}

\long\def\symbolfootnote[#1]#2{\begingroup%
\def\thefootnote{\fnsymbol{footnote}}\footnote[#1]{#2}\endgroup}

\usepackage{float}

\makeatletter
\newcommand\floatc@simplerule[2]{{\@fs@cfont #1} #2\par}
\newcommand\fs@simplerule{\def\@fs@cfont{\bfseries}\let\@fs@capt\floatc@simplerule
  \def\@fs@pre{\hrule height1.2pt depth0pt \kern4pt}%
  \def\@fs@mid{\vspace*{0.5em} \hrule height.3pt depth0pt \vspace*{0.8em} \kern4pt}%
  \def\@fs@post{\kern4pt \hrule height1.2pt depth0pt \kern4pt \relax}%
  \let\@fs@iftopcapt\iftrue}
\makeatother

\floatstyle{boxed}
\newfloat{algorithm}{thp}{alg}	
\floatname{algorithm}{Box}

\begin{document}

\clearpage

\titleman{Extending Statistical Boosting \\ \vspace{0.5cm}
 \Large An Overview of Recent Methodological Developments\symbolfootnote[1]{This article
is not an exact copy of the original published article in \emph{Methods of Information in Medicine}. The definitive publisher-authenticated version is available online (together with the companion review and an invited discussion) at: \url{http://dx.doi.org/10.3414/ME13-01-0123}. \\
If citing, please refer to the original article: \\
 Mayr A, Binder H, Gefeller O, Schmid M. Extending Statistical Boosting --  An Overview of Recent Methodological Developments. Methods Inf Med 2014; 53(6): 428--435.  }}


\begin{center}
Andreas Mayr\symbolfootnote[2]{ \textit{Address for correspondence:}
Andreas Mayr, Institut f\"ur Medizininformatik, Biometrie und
Epidemiologie, Friedrich-Alexander Universit\"at
Erlangen-N\"urnberg, Waldstr. 6, 91054 Erlangen, Germany.}$^1$,
Harald Binder$^2$, Olaf Gefeller$^1$,\\ Matthias Schmid$^{1,3}$ \vspace{0.5cm}

\begin{footnotesize}
 \footnotemark[1] Institut f\"ur Medizininformatik, Biometrie und Epidemiologie,\\
Friedrich-Alexander-Universit\"at Erlangen-N\"urnberg, Germany\\
\footnotemark[2] Institut f\"ur Medizinische Biometrie,
Epidemiologie und Informatik, \\ Johannes Gutenberg-Universit\"at Mainz, Germany \\
\footnotemark[3] Institut f\"ur Medizinische Biometrie, Informatik und Epidemiologie, \\
Rheinische Friedrich-Wilhelms-Universit\"at Bonn, Germany\\
\end{footnotesize}

\end{center}
\normalsize \vspace{1cm}

\begin{abstract}

\noindent
\textbf{Background:} Boosting algorithms to simultaneously estimate and select predictor effects in statistical models have gained substantial interest during the last decade. \\
\noindent
\textbf{Objectives:} This review highlights recent methodological developments regarding boosting algorithms for statistical modelling especially focusing on topics relevant for biomedical research.\\
\noindent
\textbf{Methods:} We suggest a unified framework for gradient boosting and likelihood-based boosting (\textit{statistical boosting}) which have been addressed strictly separated in the literature up to now. \\
\noindent
\textbf{Results:} The methodological developments on statistical boosting during the last ten years can be grouped into three different lines of research: (i) efforts to ensure variable selection leading to sparser models, (ii) developments regarding different types of predictor effects and how to choose them, (iii) approaches to extend the statistical boosting framework to new regression settings.\\
\noindent
\textbf{Conclusions:} Statistical boosting algorithms have been adapted to carry out unbiased variable selection and automated model choice during the fitting process and can nowadays be applied in almost any regression setting in combination with a large amount of different types of predictor effects.

\end{abstract}

\clearpage
\section{Introduction}

The idea of boosting emerged from the field of supervised machine learning, where the task is to construct accurate, data-driven classifiers. The concept introduced by Schapire \cite{Schapire1989} and Freund \cite{Freund90}, provided a framework to \textit{boost} the accuracy of weak classifiers (\textit{base-learners}) to become stronger ones. The introduction of AdaBoost \cite{nr:freund.schapire:1996} was the breakthrough for the success of boosting in the field of classification and machine learning.

Regarding the use of boosting in biomedical research, AdaBoost has the disadvantage that it represents what can be described as a \emph{black box}: it might yield very accurate predictions, but does not provide any insights into the structure of the relationship between the different variables. This is due to the fact that the main focus of classical supervised learning approaches lies on getting accurate predictions for a new observation $y_{\text{new}}$ of a response variable $Y$ given one or more predictors $\bm{X} = \{X_1, \dots,X_p\}$. How the prediction for $y_{\text{new}}$ is derived and how the effect of predictors can be interpreted is often not considered to be important.

In modern biomedical research, the most important application of the boosting concept lies in the area of statistical modelling (\textit{statistical boosting}, \cite{boosting_part1}). One popular model class in this context is called generalized additive models ('GAM', \cite{hastietib}) and the aim is to model the expected value of the response variable, given the observed predictors via a link function $g(\text{E}(Y|X=\bm{x})) =  f(\bm{x})$. In comparison to classical machine learning settings, the outcome variable does not necessarily be dichotomous but can also be continuous or represent count data. In cases of more than one predictor,  the function $f(\cdot)$ consists of the additive effects of the single predictors,

\begin{eqnarray*}
 f(\bm{x}) = \beta_0 +  h_1(x_1) + \cdots + h_p(x_p)
\end{eqnarray*}

where $\beta_0$ is a general intercept and the functions $h_j(\cdot)$, $j = 1, \dots p$, refer to the partial effects of the predictors $x_j$. As examples of $h_j$, one can consider a linear effect $x_j\beta_j$ or a smooth non-linear effect represented by regression splines \cite{Eilers1996}.

The aim of statistical boosting algorithms is to simultaneously estimate and select the predictor effects in $f(\cdot)$. The partial effects of the components of $X$ are fitted by separate base-learners (\textit{component-wise boosting}) which are typically simple regression-type functions like univariate linear models or penalized splines \cite{Eilers1996}. In comparison to classical applications of the boosting concept in machine learning the focus for statistical boosting is not only to generate accurate predictions, but to yield a function where the size and the shape of different predictor effects can be visualized and interpreted. This interpretability is a major advantage towards many competing machine-learning approaches such as random forests and support vector machines \cite{kruppaziegler2014, BoulesteixSchmid_comment}.

In the companion review \cite{boosting_part1}, we described the evolution of boosting algorithms from machine learning to statistical modelling. We highlighted gradient boosting \cite{friedman_2001, BuhlmannHothorn06} and likelihood-based boosting \cite{TutzBinder, TutBin2007} as two important frameworks in the context of statistical boosting. Both approaches follow a very similar structure (see Box~\ref{alg:statboost}), and they coincide in special cases, such as classical Gaussian regression, where the $L_2$ loss is used.  Statistical boosting is particularly advantageous in the presence of high-dimensional data with more candidate variables than observations: In these settings many classical estimation algorithms become unfeasible. Additionally, statistical boosting can be adapted to carry out variable selection during the model estimation and is very flexible regarding the type of predictor effects.

\begin{algorithm}[t]
\begin{enumerate}

\item[] {\bf Initialization}
    \begin{enumerate}

    \item[(1)]  Set the iteration counter $m=0$. Initialize the additive predictor $\hat{f}^{[0]}$ with a starting value e.g. $\hat{f}^{[0]} := (\bm{0})_{i=1,...,n}$. Specify a set of regression type functions as base-learners $h_1(x_1),..., h_p(x_p)$; typically each base-learner refers to one possible candidate variable.

    \end{enumerate}

\item[]{\textbf{Component-wise fitting}}

 \begin{enumerate}
   \item[(2)] Set $m := m + 1$.

   \item[(3)] Estimate the base-learners $\hat{h}_j(\cdot)$, $j = 1,\dots, p$ separately. They are either fitted to the negative gradient of the loss function evaluated at $\hat{f}^{[m-1]}$ (gradient boosting) or by penalized maximum likelihood estimation using one step of Fisher scoring with $\hat{f}^{[m-1]}$ as offset (likelihood-based boosting).

\end{enumerate}

\item[]{\textbf{Update best performing component}}

\begin{enumerate}
\item[(4)] Select the best performing component $j^*$.

\item[(5)] Update the additive predictor $\hat{f}^{[m]}$ via the corresponding base-learner  $\hat{h}_{j^*}(x_{j^*})$
\end{enumerate}

\item[]{\textbf{Iteration}}

\begin{enumerate}
\item[] Iterate steps (2) to (5) until $m=m_{\text{stop}}$. \\
\end{enumerate}

\end{enumerate}
\caption{Overview of the common structure of statistical boosting algorithms.}\label{alg:statboost}
\end{algorithm}

The main motivation for this review article is to highlight the most important methodological developments of statistical boosting during the last years. The natural starting point for such a review are the articles by B\"uhlmann and Hothorn  (2007, \cite{BuhlmannHothorn06}) and Tutz and Binder (2006, \cite{TutzBinder}). They can be considered as the breakthrough for the application of gradient boosting and likelihood-based boosting in biomedical research and evoked substantial methodological research. Following the insights gained in the companion review, we do not further distinguish between likelihood-based boosting and gradient boosting but consider a unified framework of statistical boosting. The justification for this decision is that most methodological extensions which were initially developed for either one of the two approaches could also be easily adapted to be used for the other one.

This article is structured as follows: In Section 2, we present different approaches to enhance variable selection, i.e. the selection of the most important predictors from a possibly high-dimensional set of candidate variables. In Section 3, we highlight developments which focus less on \textit{which} variables are included in the final model, but on \textit{how}: We describe newly developed base-learners to represent specific types and shapes of effects as well as approaches to let the algorithm decide which type of effect a predictor exerts on the outcome (\emph{model choice}, \cite{Kneib2009}). In Section 4 we present methodological developments aiming to extend the statistical boosting framework to new regression settings like time-to-event data, multiple parameter dimensions or discriminatory measures. We additionally describe specific adaptations of statistical boosting for the demands of high-dimensional molecular data. In Section 5 we shortly highlight some recent applications of statistical boosting in biomedical research and give an outlook on possible topics for future research.

\section{Variable selection}

One of the most important features of statistical boosting for biomedical research is its intrinsic variable selection. The selection of the most informative predictors out of a much larger set of candidate variables is of increasing relevance: in times of bio-technological advances, particularly in microarray and genome-wide association studies, there are nowadays considerably larger amounts of predictors available to select from  \cite{Fan2010}.

Variable selection in statistical boosting is a result of component-wise fitting in combination with early stopping. In component-wise boosting algorithms, each base-learner typically refers to one component of $X$ and in every boosting iteration only the best performing base-learner is selected to be included in the current additive predictor. If the algorithm is stopped before convergence (\textit{early stopping}), base-learners and the corresponding predictors which had never been selected are effectively excluded from the final model. The main tuning parameter of statistical boosting, which also controls variable selection, is therefore the stopping iteration $m_{\text{stop}}$ (Section 3.3 of the companion review \cite{boosting_part1}).


In recent boosting literature, several authors have focused on further increasing the sparsity of the resulting models. B\"uhlmann and Yu \cite{BuehlmannYu2006} stated that gradient boosting with the $L_2$ loss has a tendency to include too many predictors in the final model, especially in high-dimensional settings. They addressed this issue with an approach called \textit{sparse} boosting, in a similar version later referred to also as \textit{penalized} boosting \cite{BuehlmannChapter33}, which changes the selection criterion in step (4) of the algorithm (Box~\ref{alg:statboost}). While the original gradient boosting algorithm chooses the best performing base-learner $h_{j^*}(\cdot)$ via the least squares criterion with respect to the gradient vector,  B\"uhlmann and Yu \cite{BuehlmannYu2006} proposed applying likelihood-based information criteria (AIC, BIC or gMDL \cite{hansen2001model}): The algorithm chooses the component which optimizes the corresponding criterion at the current iteration. B\"uhlmann and Yu reported that the resulting algorithm, in combination with the $L_2$ loss, yielded sparser models than common model-based boosting without decreasing the prediction accuracy. The authors further proposed applying information criteria to determine the stopping iteration $m_{\text{stop}}$.



Information criteria, like the AIC, have often been criticized in the literature as being biased towards overfitting \cite{Hurvich89} as well as to be unstable with respect to variable selection \cite{grevenkneib}. A particular problem in the context of boosting relates to the necessity of estimating the degrees of freedom. Such estimates are (i) only available for a limited number of loss functions and (ii) potentially biased as they neglect the search of the complete predictor space in each iteration \cite{hastiecomment}. Making use of this known bias, a recently proposed sequential stopping rule for gradient boosting \cite{Mayr_mstop} combines AIC-based stopping with resampling measures. It builds up on an earlier sequential approach by Chang et al.~\cite{Chang2010} and typically yields sparser and more accurate models than the traditional AIC-based stopping.

The \textit{twin boosting} approach by B\"uhlmann and Hothorn \cite{TwinBoosting} does not depend on any information criteria and can therefore be applied in combination with any loss function. It follows the same aim of getting sparser models than common component-wise boosting but should additionally further increase the prediction accuracy. Essentially, the authors proposed applying gradient boosting twice. First, the standard component-wise gradient descent algorithm is applied with all $p$ candidate variables as possible predictors. In a second step, only the components $p^* \leq p$ that were selected in the first run are considered for another round of gradient boosting. The selection step of the algorithm was modified so that components with a high contribution to the additive predictor resulting from the first run will more likely be selected. B\"uhlmann and Hothorn showed that their proposal led to sparser models by effectively reducing the number of variables that are selected but actually have no effect on the outcome (false positives) and could also lead to an increased prediction accuracy.

The \textit{stability selection} approach by Meinshausen and B\"uhlmann \cite{Meinshausen:2008} is not specifically linked to boosting but is an approach to possibly enhance and improve any existing variable selection technique. The principal idea is to iteratively use sampling without replacement (subsampling) with sample size $\frac{n}{2}$ and carry out variable selection based on the subsamples. Finally, the average selection probability is computed for each candidate variable and only variables that pass a certain threshold $\pi_{\text{thr}}$ are added to the final model. By incorporating the threshold, the mechanism is effectively controlling the family-wise error rate concerning the occurrence of false positives in high-dimensional data settings. In a recent article, Shah and Samworth \cite{shah2013variable} proposed a variant called \emph{complementary pairs stability selection} which uses always both disjunct subsamples and provides further improved error bounds for false positives. In fact, the idea of combining variable selection with resampling techniques is not new; as Hothorn \cite{Hothorncomment} pointed out, this basic principle was already applied in combination with bootstrapping Cox regression models by Sauerbrei and Schumacher \cite{SauerbreiCox}.

The stability selection approach is very general and can be applied without the need to alter the underlying variable selection technique. A potential problem in combination with boosting is using a sample size of $\frac{n}{2}$ for subsampling. Because the variable selection properties of boosting algorithms depend on the sample size \cite{Buehlmann2006} the selection rates on the subsamples could be biased. Schmid et al.~\cite{schmidetal:sagm} proposed using a sample-size of $\frac{4}{5} \cdot n$ to avoid greater differences between the subsamples and the complete sample.

Another issue arising from the combination of boosting with stability selection concerns the tuning of the algorithm. How should the selection of the stopping iteration $m_{\text{stop}}$ be carried out, particularly when a final model is needed for predictions? For the subsamples, $m_{\text{stop}}$ can be selected applying resampling techniques. By averaging over the subsamples, stability selection then results in a set of \textit{stable} variables which have a higher selection rate than $\pi_{\text{thr}}$.  Given the primary aim of the analysis is to identify such variables, no final model is needed.  If the focus lies on identifying a small subset of important variables \textit{and}  predicting the outcome, it remains unclear how to fit the final model with stable variables. It seems reasonable to boost an additive predictor including only the stable variables with an arbitrarily high $m_{\text{stop}}$. As a result, the final model will include all the stable variables but the prediction accuracy might be worse than for any of the underlying models, as no shrinkage of effect estimates is incorporated. Furthermore, also some of the excluded unstable variables might also add to the predictive power of a model. In this context, it can be argued that high prediction accuracy and very sparse and interpretable models might be two different goals that are often hard to achieve at the same time \cite{Hothorncomment}.

\section{Type of predictor effects}

\subsection{New base-learners for specific predictor effects}

Instead of focusing on \textit{which} variables are selected, another line of research addresses the problem \textit{how} such variables are incorporated into the final model. An important development in this context is the introduction of penalized spline (P-splines,  \cite{Eilers1996}) base-learners for gradient boosting by Schmid and Hothorn \cite{Schmid:Hothorn:boosting-p-Splines}, analogously to the work of Tutz and Binder \cite{TutzBinder} for likelihood-based boosting. While B\"uhlmann and Yu \cite{BuehlmannYu2003} had focused on  smoothing splines to include the non-linear effect of predictors on the response, P-splines have computational advantages as the dimensionality of the penalty is greatly reduced.

In practice, spline base-learners are used with a fixed parameter controlling the smoothness of the effect \cite{Schmid:Hothorn:boosting-p-Splines, BuehlmannYu2003}. However, that does not mean that every spline in the final model has the same smoothness or complexity: The overall smoothness of the effect in the final model mostly depends on the selection of the stopping iteration. As the same spline base-learner can be chosen and updated in various iterations and the final solution is the sum of those base-learner effects, boosting can adapt to an arbitrarily higher-order smoothness and complexity \cite{Schmid:Hothorn:boosting-p-Splines, BuehlmannYu2003}. For an example, see the Appendix of the companion review \cite{boosting_part1}. An extension to two-dimensional P-splines to include spatial effects or interactions was proposed by Kneib et al.~\cite{Kneib2008}. For an application incorporating also spatio-temporal effects see Robinzonov and Hothorn \cite{robinzonov2010spatial}. To include discrete spatial effects such as a regional structure, Sobotka and Kneib \cite{Sobotka:2012} proposed a Markov random field base-learner applying a penalization which ensures that neighboring regions share similar effects.

In the presence of correlated or clustered response observations, as frequently encountered in longitudinal studies containing repeated measurements, it is often necessary to include \textit{random} effects in the additive model to adjust for the underlying data structure. The corresponding model class is often referred to as \textit{mixed models} and has become increasingly popular especially in biomedical research. In combination with gradient boosting, Kneib et al.~\cite{Kneib2009} proposed a ridge-penalized base-learner with adjusted degrees of freedom for subject-specific linear or categorical variables. For likelihood-based boosting, a recent article by Groll and Tutz \cite{GroTut2012} proposed \textit{random} P-splines to boost generalized additive mixed models.

Hofner et al.~\cite{Hofner_contrained} introduced a gradient boosting framework for constrained regression, proposing specifically designed base-learners to incorporate monotonic effects for splines and ordered factors. These base-learners, including also cyclic splines, are particularly useful to incorporate subject-specific knowledge about the shape of effects. For a related approach in the context of likelihood-based boosting see Leitenstorfer and Tutz \cite{leitenstorfer2007}.

\subsection{Model choice}

The question of how to specify the particular type of effect for the candidate variables (\textit{model choice}) was investigated by Kneib et al.~\cite{Kneib2009}. Typically, this includes the automated selection of predictors that, for example, have a non-linear effect on the response and should hence be included via a P-spline base-learner instead of a simple linear model. The authors proposed using a decomposition of the linear and non-linear effect of continuous predictors, where each part is assigned to different base-learners. Afterwards, the algorithm decides which base-learner is updated. By using a reparametrization, all base-learners get the same degrees of freedom to ensure an unbiased selection between linear and non-linear effects.

The topic of an unbiased selection of base-learners was also addressed by Hofner et al.~\cite{Hofner:unbiased:2011}, who specifically focused on the different complexities of base-learners including continuous or categorical variables. The authors proposed an elaborate framework based on penalized least-squares to make all base-learners comparable with respect to their degrees of flexibility. The core of this approach is to use a modified definition of degrees of freedom, proposed already in \cite{buja1989}, given by $\text{df} = \text{trace}(2\bm{S} - \bm{S}^{\top}\bm{S})\,$ where $\bm{S}$ is the hat matrix of the penalized spline (smoother matrix). Hofner et al.~provided a formal proof that their definition is preferable compared to the common df = $\text{trace}(\bm{S})$ as it reduces selection bias. To ensure an unbiased model and variable selection, all base-learners should be hence set to the same df (e.g., df = 1). For P-splines, the decomposition in linear and non-linear effect of the continuous predictor \cite{Kneib2009} should be used. For categorical predictors, ridge penalized base-learners are applied, while for ordinal variables a ridge-type penalty for the differences of parameters of adjacent categories is proposed (compare to \cite{Gertheiss2010}).

\subsection{Block-wise boosting}

The  \textit{block-wise} boosting procedure was proposed by Tutz and Gertheiss \cite{TutzGertheiss2010} in an approach to fit functional data by boosting (for a different approach see  \cite{ferraty2009}). The general idea is to modify component-wise boosting to select not only single components but simultaneously update blocks of predictors in each iteration. In the context of ordinal variables, Gertheiss et al.~\cite{Gertheissetal2010} proposed updating groups of dummy coefficients belonging to the same categorical predictor while applying penalization on adjacent categories  (similar to \cite{Hofner:unbiased:2011}).

An adapted version of block-wise boosting was also applied by Tutz and Ulbricht \cite{TutzUlbricht2008} who focused on block-wise updates of highly correlated linear predictors. Typically, component-wise boosting algorithms have the tendency to select only one variable in case of two or more highly correlated predictors. The intention of Tutz and Ulbricht was hence to force the boosting algorithm to incorporate a grouping effect, similar to the one typically observed with the elastic net approach \cite{elasticnet} that combines $L_1$ and $L_2$ penalization. The authors introduced a correlation-based penalization which forces the algorithm to estimate similar coefficients for highly correlated predictors.

\section{Extending boosting to new regression settings}
\subsection{Boosting survival data}

The third line of research focuses on extending the framework for statistical boosting to other types of regression settings. One important feature in this context is the analysis of survival data via boosting which was already  discussed in the article by B\"uhlmann and Hothorn \cite{BuhlmannHothorn06}.  One way to deal with time-to-event data is to apply the standard $L_2$ loss and additionally incorporate inverse probability of censoring weights \cite{hothorn2006survival}.  The most popular model class for survival data, the semi-parametric Cox proportional hazard model, can be fitted both by gradient boosting \cite{Ridgeway99} as well as by likelihood-based boosting \cite{BinSch2008}. A recent article additionally proposed a likelihood-based algorithm for a smooth Cox model containing also time-varying effects \cite{hofner:cfb:2012}. Statistical boosting has also been made available for fitting Fine and Gray models in the presence of competing risks \cite{BinAllSch2009}.

Schmid and Hothorn \cite{Schmid:Hothorn:AFT-boost} extended the toolbox for boosting survival data to fully parametric accelerated failure time (AFT) models. The underlying model structure $\log(Y) = \mu + \sigma \cdot W$ contains two parameters, the location $\mu$ and the scale $\sigma$. The actual survival time is $Y$ whereas $W$ is the noise variable. The idea is to fit $\mu$ conditional on $X$ by gradient boosting of the log-likelihood and simultaneously optimize the nuisance parameter $\sigma$ in each iteration. In this context, Schmid et al.~\cite{Schmid:Multidim:2010} introduced a boosting algorithm for estimating regression models with multidimensional prediction functions. Ma et al.~\cite{ma2012methods} later proposed an approach to apply sparse boosting \cite{BuehlmannYu2006} of AFT models to construct prediction rules from gene expression data obtained from multiple studies. Boosting AFT models was also investigated by Johnson and Long \cite{johnson2011survival} who introduced a rank-based estimation, while Wang and Wang \cite{wang2010buckley} proposed Buckley-James boosting for semiparametric AFT models.

\subsection{Boosting beyond the mean}

The algorithm by Schmid et al.~\cite{Schmid:Multidim:2010} can also be used in the context of modelling count data via the negative binomial distribution  incorporating an additional scale parameter, or for ordinal outcomes via proportional odds models with additional threshold parameters \cite{schmidetalenvecolstat2011}. Following these approaches, Mayr et al.~\cite{gamboostlss:2012} proposed a modified gradient boosting algorithm for generalized additive models for location, scale and shape (GAMLSS). Ordinary regression analysis assumes that the predictors only have an effect on \textit{one} parameter of the distribution -- most often the expected mean. All other distribution parameters, such as scale and shape parameters, are treated as nuisance and are assumed to be fixed. This assumption obviously reduces complexity, however, might often not be fulfilled in practice, for example, in the presence of heteroscedasticity. The GAMLSS framework \cite{rs}, specifically aims at modelling not only the expected value of the response but the complete conditional distribution by up to four distribution parameters. Besides the 'location' as in classical GAMs, those can also represent 'scale' and 'shape' parameters -- hence the extension 'LSS'. The algorithm for boosting GAMLSS \cite{gamboostlss:2012} circles through those different parameters and therefore extends the classical component-wise boosting to multiple dimensions. A popular special case is the two-parametric beta regression for bounded outcomes \cite{schmid2013beta}.

In a recent review, Kneib \cite{kneib2013beyond} identified GAMLSS together with quantile and expectile regression as the most prominent modelling strategies that go \textit{beyond mean regression}.  While GAMLSS are evidently much more flexible than classical GAMs and particularly favorable in cases of heavily skewed or kurtotic distributions, quantile and expectile regression are completely distribution-free and hence lower the risk of model misspecification. The idea is to avoid assuming a conditional distribution and focus on directly relating the predictors to a set of quantiles or expectiles. Quantile boosting was introduced by Fenske et al.~\cite{Fens2011}, while Sobotka and Kneib \cite{Sobotka:2012} extended the statistical boosting framework to expectile regression. An advantage of boosting quantiles, compared to expectiles, is the straight-forward interpretation of the resulting predictors, which can also  be used to construct non-parametric prediction intervals \cite{MayrPI}.

A fourth regression framework going far beyond standard modelling of the mean is the conditional transformation model approach by Hothorn et al.~\cite{ctm}, which can be seen as a novel semiparametric competitor for GAMLSS. The framework builds up on additive transformation functions for separate predictors expressing effects on higher moments of the response as well. Conditional transformation models can be estimated by component-wise gradient boosting via optimizing proper scoring rules \cite{ctm}.


\subsection{Boosting for molecular data}

Statistical boosting was specifically adapted for the demands of high-dimensional molecular data, such as gene expression data \cite{Schmid:Hothorn:AFT-boost, dettling}. In this kind of applications, it is often particularly interesting  to assess the added predictive value of the molecular measurements compared to established clinical predictors. Boulesteix and Hothorn \cite{BouHot2010} therefore provided a permutation test based on gradient boosting. Likelihood-based boosting of the Cox model was adapted by Binder and Schumacher \cite{BinSch2008} to include clinical predictors as mandatory, reflected by maximum (partial) likelihood-estimation, in contrast to variable selection and penalized likelihood estimation for molecular covariates.

For high-dimensional 'omics' data, there will often be biological knowledge on the relation between covariates; e.g., pathway membership of genes or mappings between covariates that correspond to miRNA and mRNA measurements. Statistical boosting was specifically adapted to incorporate such knowledge by modifying the update scheme in the course of the boosting steps.  Binder and Schumacher \cite{BinSch2009} incorporated pathway knowledge by preferentially including genes from a certain pathway after the first gene from that pathway was selected. Gade et al.~\cite{GadPorFae2011} used a similar approach for incorporating mappings between miRNA and mRNA covariates.

While statistical boosting is a very general approach, each platform for molecular measurements provides specific challenges.  Single nucleotide polymorphism (SNP) data, for example, are often incorporated in the form of covariates that take values in $\{0, 1, 2\}$, indicating the number of minor alleles present. The variance of these covariates strongly depends on the minor allele frequency. Standardization, which is typically applied before boosting, then implies no preferential treatment of SNPs with large minor allele frequency. However, the latter might be more useful for building stable models. Binder et al.~\cite{BinBenBul2013} investigated these issues in detail and provided an adapted boosting approach.

In addition to the specific coding of covariates and their variances for different molecular platforms, there are further issues, such as measurement error, that will affect most approaches used for building a multivariable regression model. For example, the specific kind of transformation of covariates in RNA-Seq data, i.e. gene expression measured by sequencing techniques, before model building is seen to severely affect identification of important genes and prediction performance \cite{ZwiFriBin2014}.

High-dimensional molecular data also pose a challenge to boosting approaches in terms of computational complexity. Some discussion of computing time is, e.g., provided in Binder et al.~\cite{BinMueSch2012}. As indicated there, boosting for 20.000 covariates, as typically found in gene expression applications, will only take minutes on current workstations. If millions of covariates are considered, model fitting by boosting can still be performed within one or two hours. Furthermore, Binder et al.~\cite{BinBenBul2013} proposed a heuristic boosting approach for decreasing computational demand in such a setting, reducing computing times to several minutes. With such a heuristic, it might be possible to deal with an even larger number of covariates, e.g., arising from whole-genome sequencing, but main memory will currently be a limiting factor.

\subsection{Boosting discriminatory measures}

The non-parametric receiver operator characteristic (ROC) technique and the corresponding area under the ROC curve (AUC) are widely used and established discriminatory measures to assess the accuracy of diagnostic tests or low-dimensional markers. For high-dimensional genomic biomarkers, which can contain thousands of genes, the question of marker selection and combination is of high relevance. To carry out biomarker selection while optimizing the AUC, Ma and Huang \cite{ma2005} introduced a gradient boosting technique that uses a smoothed version of the AUC as objective function. In the context of discriminatory measures, gradient boosting is the more flexible statistical boosting approach as it does not depend on a likelihood but needs only the gradient of the loss function.

Wang \cite{wang2011hingeboost} introduced a gradient boosting approach to optimize the hinge-loss (\textit{HingeBoost}). The hinge loss is the standard loss function for support vector machines (SVM) and its optimization is approximately equivalent to maximizing the AUC \cite{steck2007hinge}. Via HingeBoost it is also possible to incorporate unequal misclassification costs. This special feature could be of high clinical relevance as false positives and false negatives in practice often have very different consequences. Additionally, the twin boosting approach \cite{TwinBoosting} was incorporated to enhance variable selection in high-dimensional settings. In a later work, HingeBoost was further extended to multiple classes in order to classify different cancer types \cite{wang2012multi}.

Other recent contributions in the context of boosting discriminatory measures focus on the partial area under the ROC curve (pAUC, \cite{komori2010boosting}). This measure has gained popularity because often only a part of the ROC curve is relevant in practice. Typically, a range of either sensitivity or specificity is fixed while the other measure is optimized -- the standard AUC can not reflect this asymmetry. To optimize biomarker combinations with respect to the pAUC, Schmid et al.~\cite{schmidetal:sagm} proposed a component-wise gradient boosting algorithm based on a non-parametric estimator \cite{wang2011marker}.

A popular traditional discriminatory measure for the evaluation of prediction models is the concordance index ($C$-index), by Harrell et al.~\cite{harrell1}. In case of survival data, it can be used to summarize the time-dependent AUC curve for measuring predictive performance \cite{heagerty05}. Two recent articles \cite{chen2013gradient, mayr2014CI} proposed applying gradient boosting to optimize the $C$-index: Chen et al. \cite{chen2013gradient} optimized a smoothed version of Harrell's original estimator (``Harrell's C''). The authors included sigmoid functions to make the loss function differentiable (similar to \cite{ma2005}). Mayr and Schmid \cite{mayr2014CI} additionally considered an improved estimator of the $C$-index that avoids a censoring bias by incorporating inverse-probability-of-censoring weights \cite{unoC, schmidpotapov}.

\section{Conclusion}

In our review, we suggested a unified framework for gradient boosting \emph{and} likelihood-based boosting which we referred to as \textit{statistical boosting} \cite{boosting_part1}.  These two approaches have been treated strictly separate in the literature up to now, but share the same historical roots and follow the same generic structure. In this article, we documented the significant progress in methodological research on statistical boosting over the last few years. Most extensions which were initially developed for one variant can be easily adapted to fit the other one. It therefore makes sense, also for future research, to broaden the scope and bridge the methodological gap between those two schools of thought regarding statistical boosting.

Many of the methodological developments presented here were in fact motivated by biomedical applications. Besides the specific adaptations mentioned in this article, statistical boosting has been used in much more biomedical settings due to its versatile nature. This includes applications as diverse as providing alternatives to propensity score techniques \cite{ReiPorSta2013}, improving capture-recapture approaches for systematic literature reviews \cite{RueReiMot2011}, analysis of child stunting \cite{fenske2013plos} or predicting birth weight based on ultrasound images of the fetus \cite{faschingbauer2012macro}.

Statistical boosting algorithms have arguably developed into one of the most flexible estimation schemes in modern applied regression. A reason for the success story of statistical boosting, besides its methodological advantages, is surely the practical advantage of freely available software. The main algorithms are implemented in  R  add-on packages and are hence open source and freely available (see companion review \cite{boosting_part1}). Many of the extensions described in this article were either included as functions in those main packages or additionally add-on packages have been released together with the methodological paper. 

Future research might focus on extending boosting not only to multiple parameter dimensions (as described in Section~4.2) but to multiple outcome variables with one common loss function. Possible applications are the joint modelling of longitudinal and time-to-event data, as frequently encountered in clinical studies. The practical advantages of boosting regarding variable selection and model choice, its flexibility when it comes to predictor effects and the robustness in high-dimensional settings could hence be passed on to multiple additive predictors for the different outcomes. Statistical boosting could mean an attractive option for joint model estimation in practice and would further extend the methodological toolbox for regression analysis in biomedical research.

\subsection*{Acknowledgements}

The work on this article was supported by the Deutsche Forschungsgemeinschaft (DFG) (\texttt{www.dfg.de}), grant SCHM 2966/1-1. The authors thank Birgit Deutsch for her help with the literature search as well as Diana Pereira for the linguistic revision of the manuscript.

\subsection*{Discussion}

An invited discussion on this article and its companion review \cite{boosting_part1} can be found in the same issue of \textit{Methods of Information in Medicine} as the original article: \\

B\"uhlmann P, Gertheiss J, Hieke S,  Kneib T, Ma S, Schumacher M, Tutz G, Wang CY, Wang Z, Ziegler A. Discussion of ``The Evolution of Boosting Algorithms'' and ``Extending Statistical Boosting''. Methods Inf Med 2014; 53: XX-XX.

\small
\bibliography{bibliography_boosting_AM}

\begin{thebibliography}{10}

\bibitem{Schapire1989}
Schapire RE.
\newblock The Strength of Weak Learnability.
\newblock Machine Learning. 1990;5(2):197--227.

\bibitem{Freund90}
Freund Y.
\newblock Boosting a Weak Learning Algorithm by Majority.
\newblock In: Fulk MA, Case J, editors. Proceedings of the Third Annual
  Workshop on Computational Learning Theory, COLT 1990, University of
  Rochester, Rochester, NY, USA, August 6-8, 1990; 1990. p. 202--216.

\bibitem{nr:freund.schapire:1996}
Freund Y, Schapire R.
\newblock Experiments With a New Boosting Algorithm.
\newblock In: Proceedings of the Thirteenth International Conference on Machine
  Learning Theory. San Francisco, CA: San Francisco: Morgan Kaufmann Publishers
  Inc.; 1996. p. 148--156.

\bibitem{boosting_part1}
Mayr A, Binder H, Gefeller O, Schmid M.
\newblock The Evolution of Boosting Algorithms - From Machine Learning to
  Statistical Modelling.
\newblock Methods of Information in Medicine. 2014;53(6):419--427.

\bibitem{hastietib}
Hastie T, Tibshirani R.
\newblock Generalized Additive Models.
\newblock London: Chapman \& Hall; 1990.

\bibitem{Eilers1996}
Eilers PHC, Marx BD.
\newblock Flexible Smoothing with {B}-splines and Penalties (with discussion).
\newblock Statistical Science. 1996;11:89--121.

\bibitem{kruppaziegler2014}
Kruppa J, Liu Y, Biau G, Kohler M, K{\"o}nig IR, Malley JD, et~al.
\newblock Probability Estimation with Machine Learning Methods for Dichotomous
  and Multi-Category Outcome: Theory.
\newblock Biometrical Journal. 2014;To appear.

\bibitem{BoulesteixSchmid_comment}
Boulesteix AL, Schmid M.
\newblock Discussion: Machine Learning Versus Statistical Modeling.
\newblock Biometrical Journal;Available from:
  \url{http://dx.doi.org/10.1002/bimj.201300226}.

\bibitem{friedman_2001}
Friedman JH.
\newblock Greedy Function Approximation: A Gradient Boosting Machine.
\newblock The Annals of Statistics. 2001;29:1189--1232.

\bibitem{BuhlmannHothorn06}
B{\"u}hlmann P, Hothorn T.
\newblock Boosting Algorithms: Regularization, Prediction and Model Fitting
  (with Discussion).
\newblock Statistical Science. 2007;22:477--522.

\bibitem{TutzBinder}
Tutz G, Binder H.
\newblock Generalized Additive Modeling with Implicit Variable Selection by
  Likelihood-based Boosting.
\newblock Biometrics. 2006;62:961--971.

\bibitem{TutBin2007}
Tutz G, Binder H.
\newblock Boosting Ridge Regression.
\newblock Computational Statistics \& Data Analysis. 2007;51(12):6044--6059.

\bibitem{Kneib2009}
{Kneib} T, {Hothorn} T, {Tutz} G.
\newblock Variable Selection and Model Choice in Geoadditive Regression Models.
\newblock Biometrics. 2009;65(2):626--634.

\bibitem{Fan2010}
Fan J, Lv J.
\newblock A Selective Overview of Variable Selection in High Dimensional
  Feature Space.
\newblock Statistica Sinica. 2010;20:101--148.

\bibitem{BuehlmannYu2006}
B{\"u}hlmann P, Yu B.
\newblock Sparse Boosting.
\newblock Journal of Machine Learning Research. 2007;7:1001--1024.

\bibitem{BuehlmannChapter33}
B{\"u}hlmann P.
\newblock Bagging, Boosting and Ensemble Methods.
\newblock In: Gentle YJE, H{\"a}rdle W, Mori Y, editors. Handbook of
  Computational Statistics. Springer Handbooks; 2012. p. 985--1022.

\bibitem{hansen2001model}
Hansen MH, Yu B.
\newblock Model Selection and the Principle of Minimum Description Length.
\newblock Journal of the American Statistical Association.
  2001;96(454):746--774.

\bibitem{Hurvich89}
Hurvich CM, Tsai CL.
\newblock Regression and Time Series Model Selection in Small Samples.
\newblock Biometrika. 1989;76(2):237--397.

\bibitem{grevenkneib}
Greven S, Kneib T.
\newblock On the Behaviour of Marginal and Conditional Akaike Information
  Criteria in Linear Mixed Models.
\newblock Biometrika. 2010;97(4):773--789.

\bibitem{hastiecomment}
Hastie T.
\newblock Comment: Boosting Algorithms: Regularization, Prediction and Model
  Fitting.
\newblock Statistical Science. 2007;22(4):513--515.

\bibitem{Mayr_mstop}
Mayr A, Hofner B, Schmid M.
\newblock The Importance of Knowing When to Stop -- A Sequential Stopping Rule
  for Component-Wise Gradient Boosting.
\newblock Methods of Information in Medicine. 2012;51(2):178--186.

\bibitem{Chang2010}
Chang YCI, Huang Y, Huang YP.
\newblock Early Stopping in $L_2$ Boosting.
\newblock Comput Stat Data Anal. 2010;54(10):2203--2213.

\bibitem{TwinBoosting}
B\"{u}hlmann P, Hothorn T.
\newblock Twin Boosting: Improved Feature Selection and Prediction.
\newblock Statistics and Computing. 2010;20(2):119--138.

\bibitem{Meinshausen:2008}
Meinshausen N, B\"uhlmann P.
\newblock Stability Selection (with Discussion).
\newblock Journal of the Royal Statistical Society Series B. 2010;72:417--473.

\bibitem{shah2013variable}
Shah RD, Samworth RJ.
\newblock Variable Selection with Error Control: Another Look at Stability
  Selection.
\newblock Journal of the Royal Statistical Society: Series B (Statistical
  Methodology). 2013;75(1):55--80.

\bibitem{Hothorncomment}
Hothorn T.
\newblock Discussion: Stability Selection.
\newblock Journal of the Royal Statistical Society Series B. 2010;72:463--464.

\bibitem{SauerbreiCox}
Sauerbrei W, Schumacher W.
\newblock A Bootstrap Resampling Procedure for Model-Building -- Application to
  the Cox Regression-Model.
\newblock Statistics in Medicine. 1992;11:2093--2109.

\bibitem{Buehlmann2006}
B{\"u}hlmann P.
\newblock Boosting for High-Dimensional Linear Models.
\newblock The Annals of Statistics. 2006;34:559--583.

\bibitem{schmidetal:sagm}
Schmid M, Hothorn T, Krause F, Rabe C.
\newblock A {PAUC}-based Estimation Technique for Disease Classification and
  Biomarker Selection.
\newblock Statistical Applications in Genetics and Molecular Biology.
  2012;11(5).

\bibitem{Schmid:Hothorn:boosting-p-Splines}
Schmid M, Hothorn T.
\newblock Boosting Additive Models Using Component-Wise {P}-splines.
\newblock Computational Statistics \& Data Analysis. 2008;53:298--311.

\bibitem{BuehlmannYu2003}
B{\"u}hlmann P, Yu B.
\newblock Boosting with the {$L_2$} Loss: Regression and Classification.
\newblock Journal of the American Statistical Association. 2003;98:324--338.

\bibitem{Kneib2008}
Kneib T, M{\"u}ller J, Hothorn T.
\newblock Spatial Smoothing Techniques for the Assessment of Habitat
  Suitability.
\newblock Environmental and Ecological Statistics. 2008;15:343--364.

\bibitem{robinzonov2010spatial}
Robinzonov N, Hothorn T.
\newblock Boosting for Estimating Spatially Structured Additive Models.
\newblock In: Kneib T, Tutz G, editors. Statistical Modelling and Regression
  Structures. Springer; 2010. p. 181--196.

\bibitem{Sobotka:2012}
Sobotka F, Kneib T.
\newblock Geoadditive Expectile Regression.
\newblock Computational Statistics and Data Analysis. 2012;56:755--767.

\bibitem{GroTut2012}
Groll A, Tutz G.
\newblock Regularization for Generalized Additive Mixed Models by
  Likelihood-based Boosting.
\newblock Methods of Information in Medicine. 2012;51(2):168--177.

\bibitem{Hofner_contrained}
Hofner B, Kneib T, Hothorn T.
\newblock A Unified Framework of Constrained Regression.
\newblock arXiv preprint. 2014;Available from:
  \url{http://arxiv.org/abs/1403.7118}.

\bibitem{leitenstorfer2007}
Leitenstorfer F, Tutz G.
\newblock Generalized Monotonic Regression based on {B}-splines with an
  Application to Air Pollution Data.
\newblock Biostatistics. 2007;8(3):654--673.

\bibitem{Hofner:unbiased:2011}
Hofner B, Hothorn T, Kneib T, Schmid M.
\newblock A Framework for Unbiased Model Selection Based on Boosting.
\newblock Journal of Computational and Graphical Statistics. 2011;20:956--971.

\bibitem{buja1989}
Buja A, Hastie T, Tibshirani R.
\newblock Linear Smoothers and Additive Models.
\newblock The Annals of Statistics. 1989;17(2):453--510.

\bibitem{Gertheiss2010}
Gertheiss J, Tutz G.
\newblock Sparse Modeling of Categorical Explanatory Variables.
\newblock The Annals of Applied Statistics. 2010;4(4):2150--2180.

\bibitem{TutzGertheiss2010}
Tutz G, Gertheiss J.
\newblock Feature Extraction in Signal Regression: A Boosting Technique for
  Functional Data Regression.
\newblock Journal of Computational and Graphical Statistics. 2010;19:154--174.

\bibitem{ferraty2009}
Ferraty F, Vieu P.
\newblock Additive Prediction and Boosting for Functional Data.
\newblock Computational Statistics \& Data Analysis. 2009;53(4):1400--1413.

\bibitem{Gertheissetal2010}
Gertheiss J, Hogger S, Oberhauser C, Tutz G.
\newblock Selection of Ordinally Scaled Independent Variables with Applications
  to International Classification of Functioning Core Sets.
\newblock Applied Statistics. 2010;60(3):377--395.

\bibitem{TutzUlbricht2008}
Tutz G, Ulbricht J.
\newblock Penalized Regression with Correlation-Based Penalty.
\newblock Statistical Computing. 2008;19:239--253.

\bibitem{elasticnet}
Zou H, Hastie T.
\newblock Regularization and Variable Selection via the Elastic Net.
\newblock Journal of the Royal Statistical Society, Series B. 2005;67:301--320.

\bibitem{hothorn2006survival}
Hothorn T, B{\"u}hlmann P, Dudoit S, Molinaro A, Van Der~Laan MJ.
\newblock Survival Ensembles.
\newblock Biostatistics. 2006;7(3):355--373.

\bibitem{Ridgeway99}
Ridgeway G.
\newblock The State of Boosting.
\newblock Computing Science and Statistics. 1999;31:172--181.

\bibitem{BinSch2008}
Binder H, Schumacher M.
\newblock Allowing for Mandatory Covariates in Boosting Estimation of Sparse
  High-Dimensional Survival Models.
\newblock {BMC} Bioinformatics. 2008;9(14).

\bibitem{hofner:cfb:2012}
Hofner B, Hothorn T, Kneib T.
\newblock Variable Selection and Model Choice in Structured Survival Models.
\newblock Computational Statistics. 2013;28(3):1079--1101.

\bibitem{BinAllSch2009}
Binder H, Allignol A, Schumacher M, Beyersmann J.
\newblock Boosting for High-Dimensional Time-to-Event Data with Competing
  Risks.
\newblock Bioinformatics. 2009;25(7):890--896.

\bibitem{Schmid:Hothorn:AFT-boost}
Schmid M, Hothorn T.
\newblock Flexible Boosting of Accelerated Failure Time Models.
\newblock BMC Bioinformatics. 2008;9(269).

\bibitem{Schmid:Multidim:2010}
Schmid M, Potapov S, Pfahlberg A, Hothorn T.
\newblock Estimation and Regularization Techniques for Regression Models with
  Multidimensional Prediction Functions.
\newblock Statistics and Computing. 2010;20:139--150.

\bibitem{ma2012methods}
Ma S, Huang J, Xie Y, Yi N.
\newblock Identification of Breast Cancer Prognosis Markers Using Integrative
  Sparse Boosting.
\newblock Methods of Information in Medicine. 2012;51(2):152--161.

\bibitem{johnson2011survival}
Johnson BA, Long Q.
\newblock Survival Ensembles by the Sum of Pairwise Differences with
  Application to Lung Cancer Microarray Studies.
\newblock The Annals of Applied Statistics. 2011;5(2A):1081--1101.

\bibitem{wang2010buckley}
Wang Z, Wang C.
\newblock Buckley-James Boosting for Survival Analysis with High-Dimensional
  Biomarker Data.
\newblock Statistical Applications in Genetics and Molecular Biology.
  2010;9(1).

\bibitem{schmidetalenvecolstat2011}
Schmid M, Hothorn T, Maloney KO, Weller DE, Potapov S.
\newblock Geoadditive Regression Modeling of Stream Biological Condition.
\newblock Environmental and Ecological Statistics. 2011;18:709--733.

\bibitem{gamboostlss:2012}
Mayr A, Fenske N, Hofner B, Kneib T, Schmid M.
\newblock Generalized Additive Models for Location, Scale and Shape for
  High-Dimensional Data -- A Flexible Aproach Based on Boosting.
\newblock Journal of the Royal Statistical Society: Series C (Applied
  Statistics). 2012;61(3):403--427.

\bibitem{rs}
Rigby RA, Stasinopoulos DM.
\newblock Generalized Additive Models for Location, Scale and Shape (with
  discussion).
\newblock Applied Statistics. 2005;54:507--554.

\bibitem{schmid2013beta}
Schmid M, Wickler F, Maloney KO, Mitchell R, Fenske N, Mayr A.
\newblock Boosted Beta Regression.
\newblock PloS ONE. 2013;8(4):e61623.

\bibitem{kneib2013beyond}
Kneib T.
\newblock Beyond Mean Regression.
\newblock Statistical Modelling. 2013;13(4):275--303.

\bibitem{Fens2011}
{Fenske} N, {Kneib} T, {Hothorn} T.
\newblock Identifying Risk Factors for Severe Childhood Malnutrition by
  Boosting Additive Quantile Regression.
\newblock Journal of the American Statistical Association.
  2011;106(494):494--510.

\bibitem{MayrPI}
Mayr A, Hothorn T, Fenske N.
\newblock Prediction Intervals for Future {BMI} Values of Individual Children
  -- A Non-Parametric Approach by Quantile Boosting.
\newblock BMC Medical Research Methodology. 2012;12(6).

\bibitem{ctm}
Hothorn T, Kneib T, B{\"u}hlmann P.
\newblock Conditional Transformation Models.
\newblock Journal of the Royal Statistical Society: Series {B} (Statistical
  Methodology). 2014;76(1):3--27.

\bibitem{dettling}
Dettling M, B{\"u}hlmann P.
\newblock Boosting for Tumor Classification with Gene Expression Data.
\newblock Bioinformatics. 2003;19(9):1061--1069.

\bibitem{BouHot2010}
Boulesteix AL, Hothorn T.
\newblock Testing the Additional Predictive Value of High-Dimensional Molecular
  Data.
\newblock BMC Bioinformatics. 2010;11(78).

\bibitem{BinSch2009}
Binder H, Schumacher M.
\newblock Incorporating Pathway Information into Boosting Estimation of
  High-Dimensional Risk Prediction Models.
\newblock BMC Bioinformatics. 2009;10(18).

\bibitem{GadPorFae2011}
Gade S, Porzelius C, F{\"a}lth M, Brase JC, Wuttig D, Kuner R, et~al.
\newblock Graph based Fusion of {miRNA} and {mRNA} Expression Data Improves
  Clinical Outcome Prediction in Prostate Cancer.
\newblock BMC Bioinformatics. 2011;12(488).

\bibitem{BinBenBul2013}
Binder H, Benner A, Bullinger L, Schumacher M.
\newblock Tailoring Sparse Multivariable Regression Techniques for Prognostic
  Single-Nucleotide Polymorphism Signatures.
\newblock Statistics in Medicine. 2013;32(10):1778--1791.

\bibitem{ZwiFriBin2014}
Zwiener I, Frisch B, Binder H.
\newblock Transforming RNA-Seq Data to Improve the Performance of Prognostic
  Gene Signatures.
\newblock PLoS ONE. 2014 1;9(1):e85150.

\bibitem{BinMueSch2012}
Binder H, M\"{u}ller T, Schwender H, Golka K, Steffens M, Hengstler JG, et~al.
\newblock Cluster-localized sparse logistic regression for SNP data.
\newblock Statistical Applications ind Genetics and Molecular Biology.
  2012;11(4).

\bibitem{ma2005}
Ma S, Huang J.
\newblock Regularized {ROC} Method for Disease Classification and Biomarker
  Selection with Microarray Data.
\newblock Bioinformatics. 2005;21(24):4356--4362.

\bibitem{wang2011hingeboost}
Wang Z.
\newblock {HingeBoost}: {ROC}-Based Boost for Classification and Variable
  Selection.
\newblock The International Journal of Biostatistics. 2011;7(1):1--30.

\bibitem{steck2007hinge}
Steck H.
\newblock Hinge Rank Loss and the Area Under the {ROC} Curve.
\newblock In: Machine Learning: ECML 2007. Springer; 2007. p. 347--358.

\bibitem{wang2012multi}
Wang Z.
\newblock Multi-class {HingeBoost}. {M}ethod and Application to the
  Classification of Cancer Types Using Gene Expression Data.
\newblock Methods of Information in Medicine. 2012;51(2):162--167.

\bibitem{komori2010boosting}
Komori O, Eguchi S.
\newblock A Bboosting Method for Maximizing the Partial Area Under the {ROC}
  Curve.
\newblock BMC Bioinformatics. 2010;11(314).

\bibitem{wang2011marker}
Wang Z, Chang YCI.
\newblock Marker Selection via Maximizing the Partial Area Under the {ROC}
  Curve of Linear Risk Scores.
\newblock Biostatistics. 2011;12(2):369--385.

\bibitem{harrell1}
Harrell FE, Califf RM, Pryor DB, Lee KL, Rosati RA.
\newblock Evaluating the Yield of Medical Tests.
\newblock Journal of the American Medical Association. 1982;247(18):2543--2546.

\bibitem{heagerty05}
Heagerty PJ, Zheng Y.
\newblock Survival Model Predictive Accuracy and {R}{O}{C} Curves.
\newblock Biometrics. 2005;61(1):92--105.

\bibitem{chen2013gradient}
Chen Y, Jia Z, Mercola D, Xie X.
\newblock A Gradient Boosting Algorithm for Survival Analysis via Direct
  Optimization of Concordance Index.
\newblock Computational and Mathematical Methods in Medicine. 2013;2013.
\newblock Available from: \url{http://dx.doi.org/10.1155/2013/873595}.

\bibitem{mayr2014CI}
Mayr A, Schmid M.
\newblock Boosting the Concordance Index for Survival Data -- A Unified
  Framework to Derive and Evaluate Biomarker Combinations.
\newblock PloS ONE. 2014;9(1):e84483.

\bibitem{unoC}
Uno H, Cai T, Pencina MJ, D'Agostino RB, Wei LJ.
\newblock On the \uppercase{C}-statistics for Evaluating Overall Adequacy of
  Risk Prediction Procedures with Censored Survival Data.
\newblock Statistics in Medicine. 2011;30(10):1105--1117.

\bibitem{schmidpotapov}
Schmid M, Potapov S.
\newblock A Comparison of Estimators to Evaluate the Discriminatory Power of
  Time-to-Event Models.
\newblock Statistics in Medicine. 2012;31(23):2588--2609.

\bibitem{ReiPorSta2013}
Reiser V, Porzelius C, Stampf S, Schumacher M, Binder H.
\newblock Can Matching Improve the Performance of Boosting for Identifying
  Important Genes in Observational Studies?
\newblock Computational Statistics. 2013;28(1):37--49.

\bibitem{RueReiMot2011}
R{\"u}cker G, Reiser V, Motschall E, Binder H, Meerpohl JJ, Antes G, et~al.
\newblock Boosting Qualifies Capture-Recapture Methods for Estimating the
  Comprehensiveness of Literature Searches for Systematic Reviews.
\newblock Journal of Clinical Epidemiology. 2011;64(12):1364--1372.

\bibitem{fenske2013plos}
Fenske N, Burns J, Hothorn T, Rehfuess EA.
\newblock Understanding Child Stunting in India: A Comprehensive Analysis of
  Socio-Economic, Nutritional and Environmental Determinants Using Additive
  Quantile Regression.
\newblock PloS ONE. 2013;8(11):e78692.

\bibitem{faschingbauer2012macro}
Faschingbauer F, Beckmann M, Goecke T, Yazdi B, Siemer J, Schmid M, et~al.
\newblock A New Formula for Optimized Weight Estimation in Extreme Fetal
  Macrosomia ($\geq$ 4500 g).
\newblock European Journal of Ultrasound. 2012;33(05):480--488.

\end{thebibliography}

\end{document}